\documentclass[prd,aps,floats,letterpaper,floatfix,superscriptaddress,nofootinbib,preprintnumbers,twocolumn]{revtex4}
\usepackage{amsmath}
\usepackage{hyperref}
\usepackage{color}
\usepackage[mathenv]{}
\usepackage[dvipdf]{graphicx}

\newcommand{\beq}{\begin{equation}}
\newcommand{\bea}{\begin{eqnarray}}
\newcommand{\eeq}{\end{equation}}
\newcommand{\eea}{\end{eqnarray}}

\begin{document}
\title{Photothermal Fluctuations as a Fundamental Limit to Low-Frequency Squeezing\\in a Degenerate Optical Parametric Amplifier}

\author{Keisuke Goda}
\affiliation{LIGO Laboratory, Massachusetts Institute of
Technology, Cambridge, 02139, USA}

\author{Kirk McKenzie} 
\affiliation{Center for Gravitational Physics, Department of Physics, Faculty of Science, The Australian National University, ACT 0200, Australia}

\author{Eugeniy E. Mikhailov}
\affiliation{LIGO Laboratory, Massachusetts Institute of
Technology, Cambridge, 02139, USA}

\author{Ping Koy Lam} 
\affiliation{Quantum Optics Group, Department of Physics, Faculty of Science, The Australian National University, ACT 0200, Australia}

\author{David McClelland} 
\affiliation{Center for Gravitational Physics, Department of Physics, Faculty of Science, The Australian National University, ACT 0200, Australia}

\author{Nergis Mavalvala}
\affiliation{LIGO Laboratory, Massachusetts Institute of
Technology, Cambridge, 02139, USA}

\begin{abstract}
\noindent We study the effect of photothermal fluctuations on squeezed states of light through the photo-refractive effect and thermal expansion in a degenerate optical parametric amplifier (OPA). We also discuss the effect of the photothermal noise in various cases and how to minimize its undesirable consequences. We find that the photothermal noise in the OPA introduces a significant amount of noise on phase squeezed beams, making them less than ideal for low frequency applications such as gravitational wave (GW) interferometers, whereas amplitude squeezed beams are relatively immune to the photothermal noise and may represent the best choice for application in GW interferometers.   
\end{abstract}

\preprint{\large {LIGO-P050005-00-R}}
\maketitle

\section{Introduction}
Optical squeezed states are used in many areas of quantum optics to improve the sensitivity of measurements to beyond the shot noise limit (SNL). For example, squeezed states can be used in interferometers~\cite{xiao}, spatial and spectroscopic measurements~\cite{fabre,polzik}, and potentially to improve the quantum noise limit of gravitational wave (GW) interferometers~\cite{caves,kimble,mckenzie1}. Optical parametric amplifiers (OPA) are often the systems of choice to produce squeezed states since, in theory, they can produce states with very high levels of squeezing. The level of squeezing that can be produced in these systems is limited by the introduction of noise from a variety of sources. The noise sources that have been reported to limit squeezing in these systems are pump noise~\cite{wodkiewicz,crouch,zubairy} and seed noise~\cite{lam,mckenzie2}.

In experiments reported to date, the maximum amount of squeezing inferred before detection is around 7dB~\cite{lam}. This result, and in most results, the maximum squeezing is measured at sideband frequencies above 1MHz, rather than at lower frequencies where ideally, greater squeezing is predicted. There has been recent interest in producing squeezed states at lower frequencies, primarily for use in gravitational wave detectors. For such states the squeezing bandwidth should cover the GW detection band (10Hz to 10kHz). Several results have been published below 300kHz~\cite{bowen,schnabel,laurat} with the lowest result 280Hz~\cite{mckenzie2}. These frequencies represent a different regime experimentally to the majority of squeezed state production and as such other potential limiting low frequency noise sources need to be considered.

One such effect which is large at low frequencies in optical cavity systems is the photothermal-effect-induced noise~\cite{braginsky,liu,cerdonio}. This effect has not previously been reported in an OPA system to our knowledge. By nature, photothermal effects are important at low frequencies and may be significant for limiting the generation of squeezed light in the GW detection band. This effect is investigated theoretically in an OPA cavity system in this article. 

The photothermal effect can be described as the absorption of optical power in a medium causing a temperature change to the medium. This effect may be significant in most nonlinear crystals since many have relatively high absorption rates. For example, the crystal $\mbox{MgO:LiNbO}_3$ has the linear absorption rate of 4$\%$cm$^{-1}$ at 532nm~\cite{PKL}. High absorption rates coupled with the high circulating power required for strong nonlinearity result in a large amount of optical power absorbed into the crystal, which may cause a significant temperature change. The average temperature change due to the power absorbed in the crystal can be compensated for by using a temperature controller, and does not pose significant problems for most experimental systems. 

Instead, we focus our investigation on the effect of photothermally induced temperature fluctuations, caused by fluctuations in the circulating power in the OPA cavity. The circulating power fluctuations could have classical and quantum origins, or in the case of a shot-noise-limited system, only quantum mechanical origin. The photothermal noise caused by thermal-expansive noise and thermal-refractive noise has two degrading effects on the production of squeezed light in the OPA. The first effect is via fluctuations in the nonlinearity. This arises as the nonlinear strength is temperature dependent due to the phase matching condition. The second effect is via optical path length fluctuations. The temperature fluctuations cause the optical path length to change, potentially causing a detuning of the optical cavity. These effects appear as $1/(\Omega^2+\Omega_T^2)$ in variance, where $\Omega_T$ is the thermal relaxation cutoff frequency of the nonlinear medium, and are therefore primarily significant at low frequencies.

This article is organized as follows: In Section~\ref{sect:field_equations}, we write down the equations of motion for the fundamental and second-harmonic modes in an OPA with extra terms required to take into account the photothermal effect. In Section~\ref{sect:photothermal}, we quantify the fluctuating photothermal effect. In Section~\ref{sect:deltaP}, the relation between fluctuations in the power absorbed into the crystal and in its temperature is described. In Sections~\ref{sect:coupling_constant} and~\ref{sect:detunings}, the coupling of the temperature fluctuations to the fluctuations in the nonlinear coupling strength and cavity resonance frequencies through the photo-refractive effect and thermal expansion of the crystal is described. In Section~\ref{sect:field_equations_solved}, the equations of motion with these photothermal contributions are solved, and the quadrature field amplitudes in both amplitude and phase quadratures are studied. In Section~\ref{sect:results}, the squeezed/anti-squeezed quadrature variances with the inclusion of the photothermal noise are derived and plotted. In Section~\ref{sect:noise_variances}, we discuss the results for standard experimental parameters and for the shot-noise-limited case. In Section~\ref{sect:GW}, we consider the effect of squeezing with the photothermal noise on gravitational wave interferometers at low frequencies. The conclusions of the paper are summarized in Section~\ref{sect:conclusions}.

\section{Field Evolution Equations in the Degenerate Optical Parametric Amplifier}
\label{sect:field_equations}
\begin{figure}[t]
\includegraphics[width=0.45\textwidth]{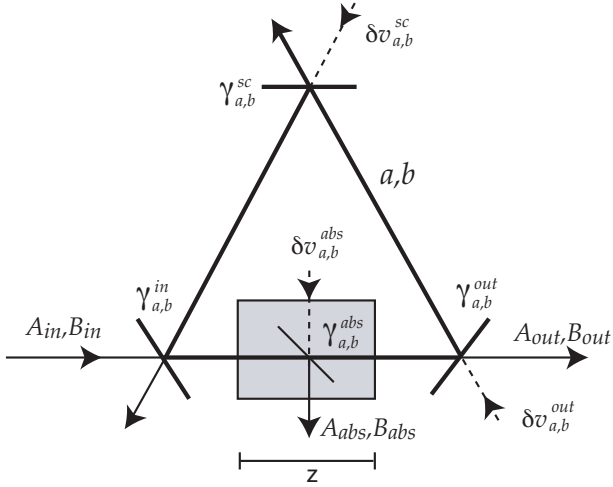}
\caption{\label{fig:fig1} A Schematic of the OPA cavity. $A_{in}$ and $B_{in}$ are the input fields to the OPA cavity, $A_{out}$ and $B_{out}$ are the output fields, and $a$ and $b$ are the intra-cavity fields at the fundamental and second-harmonic frequencies respectively. $\gamma^{in}_{a,b}$, $\gamma^{out}_{a,b}$, $\gamma^{sc}_{a,b}$, and $\gamma^{abs}_{a,b}$ are the cavity damping constants associated with the input-coupling, output-coupling, intra-cavity scattering, and intra-cavity absorption respectively. The index (a,b) refers to the fundamental and second-harmonic frequencies respectively. $\delta v^{out}_{a,b}$, $\delta v^{sc}_{a,b}$, and $\delta v^{abs}_{a,b}$ are the associated vacuum fields that couple in.}
\end{figure}
In this section, the model of the OPA based on the Heisenberg equations of motion is introduced, then these equations are linearized and additional terms for the photothermal fluctuations are introduced. This sets up the formalism to include the photothermal fluctuations which are described in terms of the input fields in the following section. The modes can then be out-coupled and the variances calculated. 

Starting from the quantum Langevin equation, the equations of motion for the intra-cavity fields at the fundamental frequency, $a$, and at the second-harmonic frequency, $b$, are given by~\cite{collett_gardiner}
\begin{eqnarray}
\label{eq:equation_of_motion_for_a}
\dot{a} &=& -\left(i\omega_a^c+\gamma_{a}^{tot}\right) a + \epsilon^* a^{\dagger}b + \sqrt{2\gamma_{a}^{in}}A_{in}e^{-i\omega_a t} \nonumber\\
& &+ \sqrt{2\gamma_{a}^{out}}\delta v_{a}^{out} + \sqrt{2\gamma_{a}^{sc}}\delta v_{a}^{sc} + \sqrt{2\gamma_{a}^{abs}}\delta v_{a}^{abs}, \\
\label{eq:equation_of_motion_for_b}
\dot{b} &=& -\left(i\omega_b^c+\gamma_{b}^{tot}\right) b -\frac{1}{2} \epsilon a^{2} + \sqrt{2\gamma_{b}^{in}}B_{in}e^{-i\omega_b t} \nonumber\\
& &+ \sqrt{2\gamma_{b}^{out}}\delta v_{b}^{out} + \sqrt{2\gamma_{b}^{sc}}\delta v_{b}^{sc} + \sqrt{2\gamma_{b}^{abs}}\delta v_{b}^{abs}.
\end{eqnarray}
The fields and coupling rates here are shown schematically in Figure~\ref{fig:fig1}. $A_{in}$ and $B_{in}$ are the fundamental and second-harmonic input fields to the cavity at frequencies $\omega_a$ and $\omega_b$ respectively ($\omega_b = 2\omega_a$). $\epsilon$ is the nonlinear coupling constant. $\omega_a^c$ and $\omega_b^c$ are the cavity resonance frequencies of the fundamental and second-harmonic fields. $\gamma^{in}_{a,b}$, $\gamma^{out}_{a,b}$, $\gamma^{sc}_{a,b}$, and $\gamma^{abs}_{a,b}$ are the cavity damping constants associated with the input-coupling, output-coupling, intra-cavity scattering, and intra-cavity absorption at both frequencies. $\delta v^{out}_{a,b}$, $\delta v^{sc}_{a,b}$, and $\delta v^{abs}_{a,b}$ are the associated vacuum fields that couple in. The following commutation relations are satisfied: 
\begin{eqnarray}
\label{eq:commutation_relations}
\left[s, s\right] = 0, \hspace{0.7cm}
\left[s, s^{\dagger}\right] = 1,
\end{eqnarray}
for $ s = A^{in}, B^{in}, \delta v_a^{out}, \delta v_b^{out}, \delta v_a^{sc}, \delta v_b^{sc}, \delta v_a^{abs}$, and $\delta v_b^{abs}$, and all others vanish.

Transforming to the rotating frame of each field with $a \rightarrow e^{i \omega_a t}a,
b \rightarrow e^{i \omega_b t}b$, and similarly for the input fields, the equations of motion become,
\begin{eqnarray}
\label{eq:equation_of_motion_for_a_rotated}
\dot{a} &=& -\left(i\omega_a^{det}+\gamma_{a}^{tot}\right) a + \epsilon^* a^{\dagger}b + \sqrt{2\gamma_{a}^{in}}A_{in} \nonumber\\
& &+ \sqrt{2\gamma_{a}^{out}}\delta v_{a}^{out} + \sqrt{2\gamma_{a}^{sc}}\delta v_{a}^{sc} + \sqrt{2\gamma_{a}^{abs}}\delta v_{a}^{abs}, \\
\label{eq:equation_of_motion_for_b_rotated}
\dot{b} &=& -\left(i\omega_b^{det}+\gamma_{b}^{tot}\right) b -\frac{1}{2} \epsilon a^{2} + \sqrt{2\gamma_{b}^{in}}B_{in} \nonumber\\
& &+ \sqrt{2\gamma_{b}^{out}}\delta v_{b}^{out} + \sqrt{2\gamma_{b}^{sc}}\delta v_{b}^{sc} + \sqrt{2\gamma_{b}^{abs}}\delta v_{b}^{abs},
\end{eqnarray}
where the cavity detunings $\omega_a^{det} = \omega_a^c - \omega_a$ and $\omega_b^{det} = \omega_b^c - \omega_b$. 

The most common method of generating the analytic form of squeezed quadrature variances is to expand the operators about their steady state values and then linearize the resulting expressions to first order in the fluctuation terms~\cite{yurke}. To linearize the equations of motion, the following substitution for the annihilation and creation operators are used:
\begin{eqnarray}
a &=& \bar{a} + \delta a, \hspace{0.7cm} a^{\dagger} = \bar{a}^* + \delta a^{\dagger}, \\
b &=& \bar{b} + \delta b, \hspace{0.7cm} b^{\dagger} = \bar{b}^* + \delta b^{\dagger}, 
\end{eqnarray}
and similarly for the input fields $A_{in},B_{in}$. $\bar{x}$ for $x=a,b$ is the complex expectation value $\left\langle x\right\rangle$ and $\delta x$ is the operator for the fluctuations of $x$ so that $\left\langle \delta x\right\rangle = 0$.
 
To obtain the equations of motion for the fluctuating components, we consider the following fluctuations due to the photothermal effect in the crystal: (1) fluctuations in the resonance frequencies of the cavity due to fluctuations in the length of the cavity (assuming that the laser frequency is perfectly stable), and (2) fluctuations in the nonlinear coupling constant due to fluctuations in the temperature of the crystal and in the length of the nonlinear region. We then make the following substitutions:
\begin{eqnarray}
\omega_a^{det} &=& \bar{\omega}_a^{det} + \delta \omega_a^{det}, \hspace{0.5cm} 
\omega_b^{det} = \bar{\omega}_b^{det} + \delta \omega_b^{det}, \\
& &\mbox{($\omega_a^{det}$ and $\omega_b^{det}$ are real)},\nonumber\\
\epsilon &=& \bar{\epsilon} + \delta \epsilon, \hspace{0.5cm}
\epsilon^* = \bar{\epsilon}^* + \delta \epsilon^{\dagger}. 
\end{eqnarray}
The fluctuation terms are obtained by taking the fluctuation components in Eqs.~\eqref{eq:equation_of_motion_for_a_rotated} and~\eqref{eq:equation_of_motion_for_b_rotated},
\begin{eqnarray}
\label{eq:equation_of_motion_for_deltaa}
\delta \dot{a} &=& i\bar{a}\delta\omega_a^{det} + \left(i\bar{\omega}_a^{det}-\gamma_{a}^{tot}\right)\delta a 
+ \bar{\epsilon}^*\bar{b}\delta a^{\dagger} \nonumber\\
&&+ \bar{\epsilon}^*\bar{a}^*\delta b + \bar{a}^*\bar{b} \delta\epsilon^{\dagger} + \sqrt{2\gamma_{a}^{in}}\delta A_{in} + \sqrt{2\gamma_{a}^{out}}\delta v_{a}^{out} \nonumber\\
&&+ \sqrt{2\gamma_{a}^{sc}}\delta v_{a}^{sc} + \sqrt{2\gamma_{a}^{abs}}\delta v_{a}^{abs},\\
\label{eq:equation_of_motion_for_deltab}
\delta \dot{b} &=& i\bar{b}\delta\omega_b^{det} + \left(i\bar{\omega}_b^{det}-\gamma_{b}^{tot}\right)\delta b - \bar{\epsilon}\bar{a}\delta a \nonumber\\ 
&&-\frac{1}{2}\bar{a}^2\delta \epsilon + \sqrt{2\gamma_{b}^{in}}\delta B_{in} + \sqrt{2\gamma_{b}^{out}}\delta v_{b}^{out} \nonumber\\
&&+ \sqrt{2\gamma_{b}^{sc}}\delta v_{b}^{sc} + \sqrt{2\gamma_{b}^{abs}}\delta v_{b}^{abs}.
\end{eqnarray}

The coherent components of the equations of motion are obtained similarly by taking the coherent terms in Eqs.~\eqref{eq:equation_of_motion_for_a_rotated} and~\eqref{eq:equation_of_motion_for_b_rotated} in the steady state, assuming that the pump field is undepleted ($\epsilon \bar{a}^2\ll \sqrt{\gamma_b^{in}}\bar{B}_{in}$), 
\begin{eqnarray}
\label{eq:coherent_components}
0&=&\left(i\bar{\omega}_a^{det}-\gamma_{a}^{tot}\right)\bar{a}+\bar{\epsilon}^* \bar{a}^*\bar{b}+\sqrt{2\gamma_{a}^{in}}\bar{A}_{in},\\
0&\cong&\left(i\bar{\omega}_b^{det}-\gamma_{b}^{tot}\right)\bar{b} +\sqrt{2\gamma_{b}^{in}}\bar{B}_{in},
\end{eqnarray}
from which we find the coherent intra-cavity field amplitudes,
\begin{eqnarray}
\label{eq:a_bar}\hspace{-0.4cm}
\bar{a}&=&\frac{\sqrt{2\gamma_{a}^{in}} \bar{A}_{in} \left[\left(i\bar{\omega}_a^{det}+\gamma_{a}^{tot}\right) + \bar{\epsilon}^* \left|\bar{b}\right| e^{i(\phi_b -2\phi_a)}\right]}{\gamma_{a}^{tot 2} + \bar{\omega}_a^{det 2} - \left|\bar{\epsilon} \right|^2\left|\bar{b}\right|^2},\\
\label{eq:b_bar}\hspace{-0.4cm}
\bar{b}&\cong&\frac{\sqrt{2\gamma_{b}^{in}}}{\gamma_{b}^{tot}-i\bar{\omega}_b^{det}}\bar{B}_{in},
\end{eqnarray}
where $\phi_a$ and $\phi_b$ are the phases of the fundamental and second-harmonic input fields such that $\bar{A}_{in} = |\bar{A}_{in}|e^{i\phi_a}$ and $\bar{B}_{in} = |\bar{B}_{in}|e^{i\phi_b}$. The relative phase of the fundamental and second-harmonic fields determines whether the fundamental field is parametrically amplified or de-amplified.

Eqs.~\eqref{eq:equation_of_motion_for_deltaa}, ~\eqref{eq:equation_of_motion_for_deltab}, and their correlated fluctuation operators can be rewritten in a compact form,
\begin{eqnarray}
\label{eq:equation_of_motion_compact}
\dot{\cal{X}}_c &=& \textbf{M}_c {\cal{X}}_c + \textbf{M}_{in} {\cal{X}}_{in} + \textbf{M}_{out} {\cal{V}}_{out} \nonumber\\
& &+ \textbf{M}_{sc} {\cal{V}}_{sc} + \textbf{M}_{abs} {\cal{V}}_{abs} + {\cal{X}}_{pt},
\end{eqnarray}
where the intra-cavity and input field vectors are defined by
\begin{eqnarray}
\label{eq:field_vectors}
{\cal{X}}_c \equiv
\left(\begin{array}{ccc}
\delta a\\
\delta a^{\dagger}\\
\delta b\\
\delta b^{\dagger}
\end{array}\right),\hspace{0.3cm}
{\cal{X}}_{in} \equiv
\left(\begin{array}{ccc}
\delta A_{in}\\
\delta A_{in}^{\dagger}\\
\delta B_{in}\\
\delta B_{in}^{\dagger}
\end{array}\right),\\\nonumber
\end{eqnarray}
the varuum field vectors associated with the output coupling, absorption loss, and scattering loss are respectively defined by
\begin{eqnarray}
{\cal{V}}_{out} &\equiv&
\left(\begin{array}{ccc}
\delta v_a^{out}\\
\delta v_a^{out \dagger}\\
\delta v_b^{out}\\
\delta v_b^{out \dagger}
\end{array}\right),\hspace{0.3cm}
{\cal{V}}_{abs} \equiv
\left(\begin{array}{ccc}
\delta v_a^{abs}\\
\delta v_a^{abs \dagger}\\
\delta v_b^{abs}\\
\delta v_b^{abs \dagger}
\end{array}\right),\nonumber\\
{\cal{V}}_{sc} &\equiv&
\left(\begin{array}{ccc}
\delta v_a^{sc}\\
\delta v_a^{sc \dagger}\\
\delta v_b^{sc}\\
\delta v_b^{sc \dagger}
\end{array}\right),
\end{eqnarray}
and the field vector due to photothermal fluctuations, ${\cal{X}}_{pt}$, is split into the fluctuating nonlinear coupling constant component and the fluctuating cavity detuning component,
\begin{eqnarray}
\label{eq:Psi_pt}
{\cal{X}}_{pt} = {\cal{X}}_{\epsilon} + {\cal{X}}_{w},
\end{eqnarray}
where
\begin{eqnarray}
{\cal{X}}_{\epsilon} 
=
\left(\begin{array}{ccc}
\bar{a}^*\bar{b}\delta {\epsilon}^{\dagger}\\
\bar{a}\bar{b}^*\delta {\epsilon}\\
-\frac{1}{2}\bar{a}^{2}\delta \epsilon\\
-\frac{1}{2}\bar{a}^{*2}\delta \epsilon^{\dagger}
\end{array}\right), \hspace{0.3cm}
{\cal{X}}_{w}
=
\left(\begin{array}{ccc}
i\bar{a}\delta w_a^{det}\\
-i\bar{a}^* \delta w_a^{det \dagger}\\
i\bar{b}\delta w_b^{det}\\
-i\bar{b}^*\delta w_b^{det \dagger}
\end{array}\right).
\end{eqnarray}
$\delta \epsilon$, $\delta \omega^{det}$, and their adjoints will be derived in the following section. The coupling matrices associated with the intra-cavity field, input coupling, output coupling, absorption, and scattering are respectively defined by
\begin{eqnarray}
\textbf{M}_c &\equiv& \nonumber\\
&&\hspace{-1.7cm}\left(\begin{array}{cccc}
i\bar{\omega}_a^{det} - \gamma_a^{tot} & \bar{\epsilon}^* \bar{b} & \bar{\epsilon}^* \bar{a}^* & 0 \\
\bar{\epsilon} \bar{b}^* & -i\omega_a^{det} - \gamma_a^{tot} & 0 & \bar{\epsilon}\bar{a} \\
-\bar{\epsilon} \bar{a} & 0 & i\omega_b^{det} - \gamma_b^{tot} & 0 \\
0 & -\bar{\epsilon}^* \bar{a}^* & 0 & -i\omega_b^{tot} - \gamma_b^{tot}
\end{array}\right),\nonumber\\
\textbf{M}_{in} &\equiv& 
\left(\begin{array}{cccc}
\sqrt{2\gamma_a^{in}} & 0 & 0 & 0\\
0 & \sqrt{2\gamma_a^{in}} & 0 & 0\\
0 & 0 & \sqrt{2\gamma_b^{in}} & 0\\
0 & 0 & 0 & \sqrt{2\gamma_b^{in}}
\end{array}\right), \nonumber\\
\textbf{M}_{out} &\equiv&
\left(\begin{array}{cccc}
\sqrt{2\gamma_a^{out}} & 0 & 0 & 0\\
0 & \sqrt{2\gamma_a^{out}} & 0 & 0\\
0 & 0 & \sqrt{2\gamma_b^{out}} & 0\\
0 & 0 & 0 & \sqrt{2\gamma_b^{out}}
\end{array}\right), \nonumber
\end{eqnarray}
\begin{eqnarray}
\textbf{M}_{abs} &\equiv&
\left(\begin{array}{cccc}
\sqrt{2\gamma_a^{abs}} & 0 & 0 & 0\\
0 & \sqrt{2\gamma_a^{abs}} & 0 & 0\\
0 & 0 & \sqrt{2\gamma_b^{abs}} & 0\\
0 & 0 & 0 & \sqrt{2\gamma_b^{abs}}
\end{array}\right), \nonumber\\
\textbf{M}_{sc} &\equiv&
\left(\begin{array}{cccc}
\sqrt{2\gamma_a^{sc}} & 0 & 0 & 0\\
0 & \sqrt{2\gamma_a^{sc}} & 0 & 0\\
0 & 0 & \sqrt{2\gamma_b^{sc}} & 0\\
0 & 0 & 0 & \sqrt{2\gamma_b^{sc}}
\end{array}\right).
\end{eqnarray}

In terms of frequency components, defined by~\cite{collett_gardiner}
\begin{eqnarray}
\tilde{Q}(\Omega) = \int^{\infty}_{-\infty}Q(t)e^{i\Omega t}dt,
\end{eqnarray}
Eq.~\eqref{eq:equation_of_motion_compact} becomes
\begin{eqnarray}
\label{eq:equations_of_motion_compact_fourier}
i\Omega\tilde{\cal{X}}_c &=& \textbf{M}_c\tilde{\cal{X}}_c + \textbf{M}_{in} \tilde{\cal{X}}_{in} + \textbf{M}_{out} \tilde{\cal{V}}_{out} \nonumber\\
& &+ \textbf{M}_{sc} \tilde{\cal{V}}_{sc} + \textbf{M}_{abs} \tilde{\cal{V}}_{abs} + \tilde{\cal{X}}_{pt},
\end{eqnarray}
where $\Omega$ is the sideband frequency relative to $\omega_a$. The last field vector $\tilde{\cal{X}}_{pt}$ will be derived in the next section. The commutation relations in Eq.~\eqref{eq:commutation_relations} imply 
\begin{eqnarray}
\label{eq:commutation_relations_fourier}
\left[\tilde{s}(\Omega), \tilde{s}(\Omega')\right] = 0, \hspace{0.3cm}
\left[\tilde{s}(\Omega), \tilde{s}^{\dagger}(\Omega')\right] = \delta(\Omega-\Omega^{'}),
\end{eqnarray}
for $ s = \delta A^{in}, \delta B^{in}, \delta v_a^{out}, \delta v_b^{out}, \delta v_a^{sc}, \delta v_b^{sc}, \delta v_a^{abs}$, and $\delta v_b^{abs}$, and the commutation relations between any two different states is zero.   

\section{Photothermal Noise}
\label{sect:photothermal}
Throughout this paper, we consider only type I phase-matching which is simple and has been shown to generate squeezing at low frequencies. The photothermal noise is described as follows: squeezing is degraded by fluctuations in (1) the nonlinear coupling strength and (2) the cavity resonance frequencies, caused by temperature fluctuations due to fluctuations in the photon power absorbed in the crystal. Since the nonlinear coupling strength is a function of the refractive index along the ordinary and extraordinary axes and the crystal length, fluctuations in the temperature of the crystal cause fluctuations in the nonlinear coupling strength. In addition, since the cavity resonance frequencies are functions of the crystal length at both the fundamental and second-harmonic frequencies, fluctuations in the crystal's temperature cause fluctuations in the resonance frequencies. In general, nonlinear crystals are absorptive and therefore, the fluctuations in the crystal's temperature are directly coupled with the fluctuations in the amplitudes of the input fields. Here we do not consider the three-dimensional expansion of the crystal. 

\subsection{Power Absorption in the Crystal}
\label{sect:deltaP}
The fluctuations in the temperature of the crystal are due to the fluctuations in the optical power absorbed into the crystal, which is directly related to the intra-cavity field fluctuations. Taking into consideration that the absorption occurs over the entire length of the crystal, the total absorbed power is given by
\begin{eqnarray}
\hspace{-0.4cm}P_{abs} &=& \bar{P}_{abs} + \delta P_{abs} \nonumber\\
&=& \hbar \omega_a A_{abs}^{\dagger} A_{abs} + \hbar \omega_b B_{abs}^{\dagger} B_{abs} \nonumber\\
&\cong& \hbar \omega_a \left[ |\bar{A}_{abs}|^2 + |\bar{A}_{abs}|(\delta A_{abs}^{\dagger} + \delta A_{abs})\right] \nonumber\\
& &+ \hbar \omega_b \left[ |\bar{B}_{abs}|^2 + |\bar{B}_{abs}|(\delta B_{abs}^{\dagger} +  \delta B_{abs})\right],
\end{eqnarray}
where the fluctuation terms are
\begin{eqnarray}
\delta A_{abs} &=& \sqrt{2\gamma_{abs}^a} \delta a - \delta v_{abs}^a, \nonumber\\
\delta A_{abs}^{\dagger} &=& \sqrt{2\gamma_{a}^{abs}} \delta a^{\dagger} - \delta v_{a}^{abs \dagger}, \\
\delta B_{abs} &=& \sqrt{2\gamma_{abs}^b} \delta b - \delta v_{abs}^b, \nonumber\\
\delta B_{abs}^{\dagger} &=& \sqrt{2\gamma_{b}^{abs}} \delta b^{\dagger} - \delta v_{b}^{abs \dagger},
\end{eqnarray}
and the coherent terms are
\begin{eqnarray}
\bar{A}_{abs} &=& \sqrt{2\gamma_{a}^{abs}} \bar{a}, \hspace{0.7cm}
\bar{A}_{abs}^* = \sqrt{2\gamma_{a}^{abs}} \bar{a}^*,\\
\bar{B}_{abs} &=& \sqrt{2\gamma_{b}^{abs}} \bar{b}, \hspace{0.7cm}
\bar{B}_{abs}^* = \sqrt{2\gamma_{b}^{abs}} \bar{b}^*.
\end{eqnarray}
Hence, we find the power fluctuation term $\delta P_{abs}$
\begin{eqnarray}
\label{eq:deltaP}
&&\hspace{-0.4cm}\delta P_{abs} \nonumber\\
&&\hspace{-0.3cm}= \hbar \omega_a |\bar{A}_{abs}|(\delta A_{abs} +  \delta A_{abs}^{\dagger}) \nonumber\\
&&\hspace{-0.2cm}+ \hbar \omega_b |\bar{B}_{abs}|( \delta B_{abs} + \delta B_{abs}^{\dagger}) \nonumber\\
&&\hspace{-0.3cm}= \hbar \omega_a \sqrt{2\gamma_a^{abs}}|\bar{a}|\left[\sqrt{2\gamma_a^{abs}}(\delta a +  \delta a^{\dagger})-(\delta v_a^{abs} + \delta v_a^{abs \dagger})\right] \nonumber\\
&&\hspace{-0.2cm}+ \hbar \omega_b \sqrt{2\gamma_b^{abs}}|\bar{b}|\left[ \sqrt{2\gamma_b^{abs}}(\delta b + \delta b^{\dagger})-(\delta v_b^{abs} + \delta v_b^{abs \dagger})\right].\nonumber\\
&&\hspace{-0.5cm}
\end{eqnarray}

Assuming that the power absorption in the crystal is uniform over the crystal length, it is directly coupled with change in the crystal's temperature through the equation~\cite{vyatchanin}
\begin{eqnarray}
C \rho V \left(\delta\dot{T}+\frac{\delta T}{\tau_T}\right) = \delta P_{abs},
\end{eqnarray}
where $\rho$ is the crystal density, $V$ is the mode volume, $C$ is the specific heat, and $\tau_T$ is the thermal relaxation time of the crystal. $\tau_T$ sets the critical frequency (adiabatic limit) for the response to the fluctuations in the optical power and is therefore given by~\cite{cerdonio,rosa}
\begin{eqnarray}
\tau_T = \frac{1}{\Omega_T} \simeq \frac{C \rho r_0^2}{\kappa},
\end{eqnarray}
where $\kappa$ is the thermal conductivity of the crystal and $r_0$ is the radius of the nonlinear interaction between the seed and pump fields, assuming they have a Gaussian transverse profile and the interaction distance is within the Rayleigh range of the fields. Here, we have assumed that the radius of the beams is much smaller than the length of the crystal and the cross section of the beams is much smaller than the cross section of the crystal. We then find the associated temperature fluctuations $\delta \tilde{T}$ in the frequency domain,
\begin{eqnarray}
\label{eq:deltaT}
\delta \tilde{T} = \frac{\delta \tilde{P}_{abs}}{(i\Omega+\Omega_T) C \rho V}.
\end{eqnarray}

\subsection{Fluctuations in the Nonlinear Coupling Strength}
\label{sect:coupling_constant}
In this section, the fluctuations in the nonlinear coupling strength are calculated for given temperature fluctuations. This result will then be used in Eq.~\eqref{eq:equations_of_motion_compact_fourier}.

The nonlinear coupling constant $\epsilon$ is a function of the phase mismatch parameter $\Delta k$ defined by $\Delta k = 2k_a-k_b$,
\begin{eqnarray}
\label{eq:epsilon}
\epsilon = \kappa_0 z e^{i\frac{\Delta k z}{2}}\mbox{sinc}{\frac{\Delta k z}{2}},
\end{eqnarray}
where $\kappa_0$ is a constant. The refractive index of a nonlinear crystal such as magnesium-oxide doped lithium-niobate $\mbox{MgO:LiNbO}_3$ is dependent on the temperature through the photo-refractive effect, which is used for achieving type I phase-matching. The temperature and wavelength dependence of the phase-matching condition for $\mbox{MgO:LiNbO}_3$ is described by the Sellmeier equation~\cite{selmeier}, which can be approximated around the optimum temperature $T_0$ at the fundamental frequency
\begin{eqnarray}
\Delta k = \xi(T-T_0),
\end{eqnarray}
where $\xi$ is a constant whose value depends on the crystal's properties, and T is the crystal's temperature. 

The fluctuations in the crystal's temperature cause the fluctuations in the nonlinear coupling strength $\delta \epsilon$ through the photo-refractive effect and thermal expansion, 
\begin{eqnarray}
\label{eq:delta_e}
\delta \epsilon &=& \frac{\partial\epsilon}{\partial \Delta k}\delta\Delta k +\frac{\partial \epsilon}{\partial z} \delta z \nonumber\\
&=& \left(\frac{\partial\epsilon}{\partial \Delta k}\frac{d \Delta k}{dT} +\frac{\partial \epsilon}{\partial z}\frac{dz}{dT}\right) \delta T \nonumber\\
&=& \left[\left(\frac{\partial\epsilon}{\partial \Delta k}\right)\xi +\left(\frac{\partial \epsilon}{\partial z}\right)\alpha z\right] \delta T,
\end{eqnarray}
Here we have used the Selmeier equation and the thermal expansion equation,
\begin{eqnarray}
\frac{d\Delta k}{dT} &=& \xi,\\
\frac{dz}{dT} &=& \alpha z,
\end{eqnarray}
where $\alpha$ is the linear thermal expansion coefficient. From Eq.~\eqref{eq:epsilon} we obtain
\begin{eqnarray}
\frac{\partial\epsilon}{\partial\Delta k} &=& \epsilon\left(\frac{iz}{2}-\frac{1}{\Delta k}+\frac{z}{2}\cot{\frac{\Delta k z}{2}}\right),\\
\frac{\partial\epsilon}{\partial z} &=& \epsilon\frac{z}{2}\left(i+\cot{\frac{\Delta k z}{2}}\right).
\end{eqnarray}

Substituting Eq.~\eqref{eq:deltaT} into Eq.~\eqref{eq:delta_e}, we express the effect of the fluctuating nonlinear coupling constant in terms of ${\cal{X}}_c$ and ${\cal{V}}_{abs}$,
\begin{eqnarray}
\label{eq:Psi_epsilon}
\tilde{\cal{X}}_{\epsilon} = 
\textbf{M}_{\epsilon}^{c}\tilde{\cal{X}}_c + \textbf{M}_{\epsilon}^{abs}\tilde{\cal{V}}_{abs},
\end{eqnarray}
where
\widetext
\begin{eqnarray}
\textbf{M}_{\epsilon}^{c} &\equiv&
\frac{1}{i\Omega+\Omega_T}\left(\begin{array}{cccc}
\bar{a}^*\bar{b}C_a^* \sqrt{2\gamma_a^{abs}} & 
\bar{a}^*\bar{b}C_a^* \sqrt{2\gamma_a^{abs}} & 
\bar{a}^*\bar{b}C_b^* \sqrt{2\gamma_b^{abs}} & 
\bar{a}^*\bar{b}C_b^* \sqrt{2\gamma_b^{abs}} \\
\bar{a}\bar{b}^*C_a \sqrt{2\gamma_a^{abs}} & 
\bar{a}\bar{b}^*C_a \sqrt{2\gamma_a^{abs}} & 
\bar{a}\bar{b}^*C_b \sqrt{2\gamma_b^{abs}} & 
\bar{a}\bar{b}^*C_b \sqrt{2\gamma_b^{abs}} \\
-\frac{1}{2}\bar{a}^{2}C_a \sqrt{2\gamma_a^{abs}} & -\frac{1}{2}\bar{a}^{2}C_a \sqrt{2\gamma_a^{abs}} & -\frac{1}{2}\bar{a}^{2}C_b \sqrt{2\gamma_b^{abs}} & -\frac{1}{2}\bar{a}^{2}C_b \sqrt{2\gamma_b^{abs}} \\
-\frac{1}{2}\bar{a}^{*2}C_a^* \sqrt{2\gamma_a^{abs}} & -\frac{1}{2}\bar{a}^{*2}C_a^* \sqrt{2\gamma_a^{abs}} & -\frac{1}{2}\bar{a}^{*2}C_b^* \sqrt{2\gamma_b^{abs}} & -\frac{1}{2}\bar{a}^{*2}C_b^* \sqrt{2\gamma_b^{abs}}
\end{array}\right),\\
\textbf{M}_{\epsilon}^{abs} &\equiv&
\frac{1}{i\Omega+\Omega_T}\left(\begin{array}{cccc}
-\bar{a}^*\bar{b}C_a^* & 
-\bar{a}^*\bar{b}C_a^* & 
-\bar{a}^*\bar{b}C_b^* & 
-\bar{a}^*\bar{b}C_b^* \\
-\bar{a}\bar{b}^*C_a & 
-\bar{a}\bar{b}^*C_a & 
-\bar{a}\bar{b}^*C_b & 
-\bar{a}\bar{b}^*C_b \\
\frac{1}{2}\bar{a}^{2}C_a & 
\frac{1}{2}\bar{a}^{2}C_a & 
\frac{1}{2}\bar{a}^{2}C_b & 
\frac{1}{2}\bar{a}^{2}C_b \\
\frac{1}{2}\bar{a}^{*2}C_a^* & 
\frac{1}{2}\bar{a}^{*2}C_a^* & 
\frac{1}{2}\bar{a}^{*2}C_b^* & 
\frac{1}{2}\bar{a}^{*2}C_b^* 
\end{array}\right),
\end{eqnarray}
\endwidetext
and
\begin{eqnarray}
C_a &=& \frac{\hbar \omega_a \sqrt{2\gamma_{a}^{abs}}|\bar{a}|}{C\rho V}
\left[\left(\frac{\partial\epsilon}{\partial \Delta k}\right)\xi +\left(\frac{\partial \epsilon}{\partial z}\right)\alpha z\right], \\
C_b &=& \frac{\hbar \omega_b \sqrt{2\gamma_{b}^{abs}}|\bar{b}|}{C\rho V}
\left[\left(\frac{\partial\epsilon}{\partial \Delta k}\right)\xi +\left(\frac{\partial \epsilon}{\partial z}\right)\alpha z\right].
\end{eqnarray}

\subsection{Fluctuations in the Cavity Detunings}
\label{sect:detunings}
In this section, the fluctuations in the optical path length are calculated for given temperature fluctuations. This result will then be used in Eq.~\eqref{eq:equations_of_motion_compact_fourier}.

The photothermal fluctuations couple to the fluctuations in the optical path length from the following two mechanisms: the photo-refractive effect and thermal expansion. The fluctuations in the optical path length can be converted into cavity resonance frequency fluctuations, using~\cite{siegman} 
\begin{eqnarray}
\label{eq:delta_omega}
\delta \omega_a^c &=&
-\frac{2\pi c}{\lambda_a} \left(\frac{1}{n_a}\frac{dn_a}{dT}+\alpha_a\right)\delta T, \\
\delta \omega_b^c &=&
-\frac{2\pi c}{\lambda_b} \left(\frac{1}{n_b}\frac{dn_b}{dT}+\alpha_b\right)\delta T,
\end{eqnarray}
and therefore, assuming that the laser frequencies are stable, the cavity detuning fluctuations are $\delta \omega_a^{det} = \delta \omega_a^c$ and $\delta \omega_b^{det} = \delta \omega_b^c$.

Substituting Eq.~\eqref{eq:deltaT} into Eq.~\eqref{eq:delta_omega}, we similarly write the effect of the fluctuating cavity detuning component in terms of ${\cal{X}}_c$ and ${\cal{V}}_{abs}$,
\begin{eqnarray}
\label{eq:Psi_omega}
\tilde{\cal{X}}_{\epsilon} = 
\textbf{M}_{\omega}^{c}\tilde{\cal{X}}_c + \textbf{M}_{\omega}^{abs}\tilde{\cal{V}}_{abs},
\end{eqnarray}
where
\widetext
\begin{eqnarray}
\textbf{M}_{\omega}^{c} &\equiv&
\frac{1}{i\Omega+\Omega_T}\left(\begin{array}{cccc}
i\bar{a}K_a\Pi_a \sqrt{2\gamma_a^{abs}} & 
i\bar{a}K_a\Pi_a \sqrt{2\gamma_a^{abs}} & 
i\bar{a}K_a\Pi_b \sqrt{2\gamma_b^{abs}} & 
i\bar{a}K_a\Pi_b \sqrt{2\gamma_b^{abs}} \\
-i\bar{a}^*K_a\Pi_a \sqrt{2\gamma_a^{abs}} & 
-i\bar{a}^*K_a\Pi_a \sqrt{2\gamma_a^{abs}} & 
-i\bar{a}^*K_a\Pi_b \sqrt{2\gamma_b^{abs}} & 
-i\bar{a}^*K_a\Pi_b \sqrt{2\gamma_b^{abs}} \\
i\bar{b}K_b\Pi_a \sqrt{2\gamma_a^{abs}} & 
i\bar{b}K_b\Pi_a \sqrt{2\gamma_a^{abs}} & 
i\bar{b}K_b\Pi_b \sqrt{2\gamma_b^{abs}} & 
i\bar{b}K_b\Pi_b \sqrt{2\gamma_b^{abs}} \\
-i\bar{b}^*K_b\Pi_a \sqrt{2\gamma_a^{abs}} & 
-i\bar{b}^*K_b\Pi_a \sqrt{2\gamma_a^{abs}} & 
-i\bar{b}^*K_b\Pi_b \sqrt{2\gamma_b^{abs}} & 
-i\bar{b}^*K_b\Pi_b \sqrt{2\gamma_b^{abs}} 
\end{array}\right),\\
\textbf{M}_{\omega}^{abs} &\equiv& 
\frac{1}{i\Omega+\Omega_T}\left(\begin{array}{cccc}
-i\bar{a}K_a\Pi_a  & 
-i\bar{a}K_a\Pi_a  & 
-i\bar{a}K_a\Pi_b  & 
-i\bar{a}K_a\Pi_b  \\
i\bar{a}^*K_a\Pi_a  & 
i\bar{a}^*K_a\Pi_a  & 
i\bar{a}^*K_a\Pi_b  & 
i\bar{a}^*K_a\Pi_b  \\
-i\bar{b}K_b\Pi_a  & 
-i\bar{b}K_b\Pi_a  & 
-i\bar{b}K_b\Pi_b  & 
-i\bar{b}K_b\Pi_b  \\
i\bar{b}^*K_b\Pi_a  & 
i\bar{b}^*K_b\Pi_a  & 
i\bar{b}^*K_b\Pi_b  & 
i\bar{b}^*K_b\Pi_b  
\end{array}\right),
\end{eqnarray}
\endwidetext
where
\begin{eqnarray}
\Pi_a &=& \frac{\hbar \omega_a \sqrt{2\gamma_{a}^{abs}}|\bar{a}|}{C\rho V}, \\
\Pi_b &=& \frac{\hbar \omega_b \sqrt{2\gamma_{b}^{abs}}|\bar{b}|}{C\rho V},\\
K_a &=& \frac{2\pi c}{\lambda_a}\left(\frac{1}{n_a}\frac{dn_a}{dT}+\alpha_a\right),\\
K_b &=& \frac{2\pi c}{\lambda_b}\left(\frac{1}{n_b}\frac{dn_b}{dT}+\alpha_b\right).
\end{eqnarray}

\section{Quadrature Field Amplitudes with the Photothermal Noise}
\label{sect:field_equations_solved}
Now that we have described the equations of motion, the absorbed power fluctuations, and the associated fluctuations in the nonlinear coupling strength and cavity detunings, we are in a position to put these equations together and solve the equations of motion with the photothermal effect. We also discuss a limiting case in which the quadrature variances can be approximated to simple analytic forms under realistic assumptions.

Substituting Eqs.~\eqref{eq:Psi_epsilon} and~\eqref{eq:Psi_omega} into Eq.~\eqref{eq:Psi_pt} yields 
\begin{eqnarray}
\label{eq:Psi_pt_total}
\tilde{\cal{X}}_{pt} = 
(\textbf{M}_{\epsilon}^{c} + \textbf{M}_{\omega}^{c}) \tilde{\cal{X}}_{c}
+
(\textbf{M}_{\epsilon}^{abs} + \textbf{M}_{\omega}^{abs}) 
\tilde{\cal{V}}_{abs},
\end{eqnarray}
and substituting this into Eq.~\eqref{eq:equations_of_motion_compact_fourier} gives
\begin{eqnarray}
\label{eq:equations_of_motion_solved}
&&\hspace{-0.7cm}(i\Omega \textbf{I}-\textbf{M}_c-\textbf{M}_{\epsilon}^{c}-\textbf{M}_{\omega}^{c}) \tilde{\cal{X}}_c \nonumber\\
&&\hspace{0.1cm}=
\textbf{M}_{in} \tilde{\cal{X}}_{in} 
+ \textbf{M}_{out} \tilde{\cal{V}}_{out} 
+ \textbf{M}_{sc} \tilde{\cal{V}}_{sc} \nonumber\\
&&\hspace{0.4cm} + (\textbf{M}_{abs} + \textbf{M}_{\epsilon}^{abs} + \textbf{M}_{\omega}^{abs}) \tilde{\cal{V}}_{abs},
\end{eqnarray}
where $\textbf{I}$ is the identity matrix. We thus find the intra-cavity field fluctuations
\begin{eqnarray}
\label{eq:intracavity_field_fluctuations}
\tilde{\cal{X}}_c &=& (i\Omega \textbf{I}-\textbf{M}_c-\textbf{M}_{\epsilon}^{c}-\textbf{M}_{\omega}^{c})^{-1}\left[\textbf{M}_{in} \tilde{\cal{X}}_{in} + \textbf{M}_{out} \tilde{\cal{V}}_{out} \right.\nonumber\\
&&\left.+ \textbf{M}_{sc} \tilde{\cal{V}}_{sc} + (\textbf{M}_{abs} + \textbf{M}_{\epsilon}^{abs} + \textbf{M}_{\omega}^{abs}) \tilde{\cal{V}}_{abs}\right].
\end{eqnarray}
Defining the extra-cavity field vector by~\cite{collett_gardiner}
\begin{eqnarray}
\label{eq:extracavity_field_fluctuations}
\tilde{\cal{X}}_{out} 
\equiv
\left(\begin {array}{cccc}
\delta \tilde{A}_{out}\\
\delta \tilde{A}_{out}^{\dagger}\\
\delta \tilde{B}_{out}\\
\delta \tilde{B}_{out}^{\dagger}
\end{array}\right),
\end{eqnarray}
we find
\widetext
\begin{eqnarray}
\hspace{-0.5cm}\tilde{\cal{X}}_{out} 
&=&  \textbf{M}_{out}\tilde{\cal{X}}_{c} - \tilde{\cal{V}}_{out} \nonumber\\
&=&
\textbf{M}_{out} 
(i\Omega \textbf{I} -\textbf{M}_c -\textbf{M}_{\epsilon}^{c} -\textbf{M}_{\omega}^{c} )^{-1} \textbf{M}_{in} \tilde{\cal{X}}_{in} + \left[\textbf{M}_{out}(i\Omega \textbf{I} -\textbf{M}_c -\textbf{M}_{\epsilon}^{c} -\textbf{M}_{\omega}^{c} )^{-1}\textbf{M}_{out} -\textbf{I}\right]\tilde{\cal{V}}_{out} \nonumber\\
& &\hspace{-0.25cm}+ \textbf{M}_{out}(i\Omega \textbf{I} -\textbf{M}_c -\textbf{M}_{\epsilon}^{c} -\textbf{M}_{\omega}^{c} )^{-1}\textbf{M}_{sc} \tilde{\cal{V}}_{sc} + \textbf{M}_{out}(i\Omega \textbf{I} -\textbf{M}_c -\textbf{M}_{\epsilon}^{c} -\textbf{M}_{\omega}^{c} )^{-1}(\textbf{M}_{abs} + \textbf{M}_{\epsilon}^{abs} + \textbf{M}_{\omega}^{abs}) \tilde{\cal{V}}_{abs}.
\end{eqnarray}
\endwidetext
It is important to note that as the photothermal effect is turned off by setting $\sqrt{\gamma_a^{abs}} \rightarrow 0$ and $\sqrt{\gamma_b^{abs}} \rightarrow 0$, the photothermal coupling matrices $M_{\epsilon}^{c}$, $M_{\epsilon}^{abs}$, $M_{\omega}^{c}$, and $M_{\omega}^{abs}$ as well as $\textbf{M}_{abs}$ all become zero, and then Eq.~\eqref{eq:extracavity_field_fluctuations} reduces to the solutions of the field evolution equations without the photothermal effect,
\begin{eqnarray}
\label{eq:extracavity_field_fluctuations_reduced}
\tilde{\cal{X}}_{out} &=&  
\textbf{M}_{out} 
\left(i\Omega \textbf{I} -\textbf{M}_c \right)^{-1} \textbf{M}_{in} \tilde{\cal{X}}_{in} \nonumber\\
& &+ \left[\textbf{M}_{out}\left(i\Omega \textbf{I} -\textbf{M}_c \right)^{-1}\textbf{M}_{out} -\textbf{I}\right]\tilde{\cal{V}}_{out} \\
& &+ \textbf{M}_{out}\left(i\Omega \textbf{I} -\textbf{M}_c \right)^{-1}\textbf{M}_{sc} \tilde{\cal{V}}_{sc}.
\end{eqnarray}

We define the amplitude and phase quadrature field fluctuation amplitudes in the frequency domain relative to the fundamental frequency respectively
\begin{eqnarray}
\delta \tilde{X}_s^1(\Omega) &\equiv& \delta \tilde{s}(\omega_a + \Omega) + \delta \tilde{s}^{\dagger}(\omega_a - \Omega),\\
\delta \tilde{X}_s^2(\Omega) &\equiv& i\left(\delta \tilde{s}(\omega_a + \Omega) - \delta \tilde{s}^{\dagger}(\omega_a - \Omega)\right),
\end{eqnarray}
for $s = A_{in}, A_{out}, B_{in}, B_{out}, v_a^{out}, v_b^{out}, v_a^{sc}, v_b^{sc}, v_a^{abs}$, and $v_b^{abs}$. The commutation relations~\eqref{eq:commutation_relations_fourier} imply the following values for the commutators of the quadrature field amplitudes and their adjoints:
\begin{eqnarray}
\label{eq:commutation_relations_for_field_amplitudes}
\left[\delta \tilde{X}^{1}_{s}, \delta \tilde{X}^{2\dagger}_{s'}\right] = -\left[\delta \tilde{X}^{2}_{s}, \delta \tilde{X}^{1\dagger}_{s'}\right] = -2i\delta(\Omega - \Omega')
\end{eqnarray}
for $s = A_{in}, A_{out}, B_{in}, B_{out}, v_a^{out}, v_b^{out}, v_a^{sc}, v_b^{sc}, v_a^{abs}$, and $v_b^{abs}$, and all others vanish. 

It is convenient to express Eq.~\eqref{eq:field_vectors} in terms of the quadrature field amplitudes,
\begin{eqnarray}
\tilde{\cal{X}}_{out}^{\star} &=& \Lambda \tilde{\cal{X}}_{out}, \hspace{0.3cm}
\tilde{\cal{X}}_{in}^{\star} = \Lambda \tilde{\cal{X}}_{in},  \hspace{0.3cm}
\tilde{\cal{V}}_{out}^{\star} = \Lambda \tilde{\cal{V}}_{out},  \nonumber\\
\tilde{\cal{V}}_{sc}^{\star} &=& \Lambda \tilde{\cal{V}}_{sc},  \hspace{0.3cm}
\tilde{\cal{V}}_{abs}^{\star} = \Lambda \tilde{\cal{V}}_{abs},  
\end{eqnarray}
where
\begin{eqnarray}
\Lambda 
&\equiv&
\left(\begin{array}{cccc}
1 & 1 & 0 & 0\\
i & -i & 0 & 0\\
0 & 0 & 1 & 1\\
0 & 0 & i & -i 
\end{array}\right), 
\end{eqnarray}
and 
\begin{eqnarray}
\tilde{\cal{X}}_{out}^{\star} &\equiv&
\left(\begin{array}{ccc}
\delta \tilde{X}^{1}_{A_{out}} \\
\delta \tilde{X}^{2}_{A_{out}} \\
\delta \tilde{X}^{1}_{B_{out}} \\
\delta \tilde{X}^{2}_{B_{out}} 
\end{array}\right), \hspace{0.2cm}
\tilde{\cal{X}}_{in}^{\star} \equiv
\left(\begin{array}{ccc}
\delta \tilde{X}^{1}_{A_{in}} \\
\delta \tilde{X}^{2}_{A_{in}} \\
\delta \tilde{X}^{1}_{B_{in}} \\
\delta \tilde{X}^{2}_{B_{in}} 
\end{array}\right), \nonumber\\
\tilde{\cal{V}}_{out}^{\star} &\equiv&
\left(\begin{array}{ccc}
\delta \tilde{X}^{1}_{v_a^{out}} \\
\delta \tilde{X}^{2}_{v_a^{out}} \\
\delta \tilde{X}^{1}_{v_b^{out}} \\
\delta \tilde{X}^{2}_{v_b^{out}} 
\end{array}\right), \hspace{0.2cm}
\tilde{\cal{V}}_{sc}^{\star} \equiv
\left(\begin{array}{ccc}
\delta \tilde{X}^{1}_{v_a^{sc}} \\
\delta \tilde{X}^{2}_{v_a^{sc}} \\
\delta \tilde{X}^{1}_{v_b^{sc}} \\
\delta \tilde{X}^{2}_{v_b^{sc}}
\end{array}\right), \nonumber\\
\tilde{\cal{V}}_{abs}^{\star} &\equiv&
\left(\begin{array}{ccc}
\delta \tilde{X}^{1}_{v_a^{abs}} \\
\delta \tilde{X}^{2}_{v_a^{abs}} \\
\delta \tilde{X}^{1}_{v_b^{abs}} \\
\delta \tilde{X}^{2}_{v_b^{abs}}
\end{array}\right).
\end{eqnarray}
Eq.~\eqref{eq:extracavity_field_fluctuations} can be rewritten as
\begin{eqnarray}
\tilde{\cal{X}}_{out}^{\star} = 
\Theta_{in}\tilde{\cal{X}}_{in}^{\star} + 
\Theta_{out}\tilde{\cal{V}}_{out}^{\star} + 
\Theta_{sc}\tilde{\cal{V}}_{sc}^{\star} + 
\Theta_{abs}\tilde{\cal{V}}_{abs}^{\star},
\end{eqnarray}
where the quadrature field coupling matrices are defined by
\widetext
\begin{eqnarray}
\Theta_{in} &\equiv& \Lambda\textbf{M}_{out} (i\Omega \textbf{I} -\textbf{M}_c -\textbf{M}_{\epsilon}^{c} -\textbf{M}_{\omega}^{c} )^{-1} \textbf{M}_{in} \Lambda^{-1}, \nonumber\\
\Theta_{out} &\equiv& \Lambda\left[\textbf{M}_{out}(i\Omega \textbf{I} -\textbf{M}_c -\textbf{M}_{\epsilon}^{c} -\textbf{M}_{\omega}^{c} )^{-1}\textbf{M}_{out} -\textbf{I}\right]\Lambda^{-1}, \nonumber\\
\Theta_{sc} &\equiv& \Lambda\textbf{M}_{out}(i\Omega \textbf{I} -\textbf{M}_c -\textbf{M}_{\epsilon}^{c} -\textbf{M}_{\omega}^{c} )^{-1}\textbf{M}_{sc} \Lambda^{-1},\nonumber\\
\Theta_{abs} &\equiv& \Lambda \textbf{M}_{out}(i\Omega \textbf{I} -\textbf{M}_c -\textbf{M}_{\epsilon}^{c} -\textbf{M}_{\omega}^{c} )^{-1}(\textbf{M}_{abs} + \textbf{M}_{\epsilon}^{abs} + \textbf{M}_{\omega}^{abs}) \Lambda^{-1}.
\end{eqnarray}
\endwidetext

Normalized amplitude and phase quadrature variances are given by~\cite{collett_gardiner}
\begin{eqnarray}
\tilde{V}^1_{s}(\Omega) = \left\langle |\delta\tilde{X}^1_{s}(\Omega)|^2\right\rangle, \hspace{0.3cm}
\tilde{V}^2_{s}(\Omega) = \left\langle |\delta\tilde{X}^2_{s}(\Omega)|^2\right\rangle,
\end{eqnarray}
for $s = A_{out}, B_{out}, A_{in}, B_{in}, v_a^{out}, v_b^{out}, v_a^{sc}, v_b^{sc}, v_a^{abs}$, and $v_b^{abs}$ respectively. The normalized quadrature variances of the fundamental output field can be written as a linear combination of $\tilde{V}_{A_{in}}^{1,2}, \tilde{V}_{B_{in}}^{1,2}, \tilde{V}_{v_a^{out}}^{1,2}, \tilde{V}_{v_b^{out}}^{1,2}, \tilde{V}_{v_a^{sc}}^{1,2}, \tilde{V}_{v_b^{sc}}^{1,2}, \tilde{V}_{v_a^{abs}}^{1,2}$, and $\tilde{V}_{v_b^{abs}}^{1,2}$~\cite{white}. Since the vacuum fields that couple in at the optical losses are in the minimum uncertainty state, 
\begin{eqnarray}
\tilde{V}_{v_a^{out}}^{1,2} = 
\tilde{V}_{v_b^{out}}^{1,2} = 
\tilde{V}_{v_a^{sc}}^{1,2} = 
\tilde{V}_{v_b^{sc}}^{1,2} = 
\tilde{V}_{v_a^{abs}}^{1,2} = 
\tilde{V}_{v_b^{abs}}^{1,2} = 1.
\end{eqnarray}
Therefore, we find the normalized amplitude and phase quadrature variances of the fundamental output field respectively,
\begin{eqnarray}
\label{eq:V1_Aout}
&&\hspace{-0.7cm}\tilde{V}_{A_{out}}^1(\Omega) \nonumber\\
&&\hspace{-0.6cm}= 
|\Theta^{(11)}_{in}(\Omega)|^2\tilde{V}_{A_{in}}^1(\Omega) + 
|\Theta^{(12)}_{in}(\Omega)|^2\tilde{V}_{A_{in}}^2(\Omega)\nonumber\\
&&\hspace{-0.4cm}+ 
|\Theta^{(13)}_{in}(\Omega)|^2\tilde{V}_{B_{in}}^1(\Omega) + 
|\Theta^{(14)}_{in}(\Omega)|^2\tilde{V}_{B_{in}}^2(\Omega) \nonumber\\ 
&&\hspace{-0.4cm}+ \sum^{4}_{j=1}\left(|\Theta^{(1j)}_{out}(\Omega)|^2 + |\Theta^{(1j)}_{sc}(\Omega)|^2 + |\Theta^{(1j)}_{abs}(\Omega)|^2\right),
\end{eqnarray}
\begin{eqnarray}
\label{eq:V2_Aout}
&&\hspace{-0.7cm}\tilde{V}_{A_{out}}^2(\Omega) \nonumber\\
&&\hspace{-0.6cm}=
|\Theta^{(21)}_{in}(\Omega)|^2\tilde{V}_{A_{in}}^1(\Omega) + 
|\Theta^{(22)}_{in}(\Omega)|^2\tilde{V}_{A_{in}}^2(\Omega) \nonumber\\
&&\hspace{-0.4cm}+ 
|\Theta^{(23)}_{in}(\Omega)|^2\tilde{V}_{B_{in}}^1(\Omega) + 
|\Theta^{(24)}_{in}(\Omega)|^2\tilde{V}_{B_{in}}^2(\Omega) \nonumber\\ 
&&\hspace{-0.4cm}+ \sum^{4}_{j=1}\left(|\Theta^{(2j)}_{out}(\Omega)|^2 + |\Theta^{(2j)}_{sc}(\Omega)|^2 + |\Theta^{(2j)}_{abs}(\Omega)|^2\right),
\end{eqnarray}
where the superscripts (ij) of $\Theta$'s denote the matrix elements. $\Theta^{(11)}_{in}/\Theta^{(13)}_{in}$ is the amplitude noise coupling constant of the seed/pump field, $\Theta^{(12)}_{in}/\Theta^{(14)}_{in}$ is the phase noise coupling constant of the seed/pump field, and the rest of the $\Theta$'s are the amplitude and phase noise coupling constants of the vacuum fields at the fundamental and second-harmonic frequencies. Note that the normalized quadrature variances are completely characterized by the normalized quadrature variances of the two input and vacuum fields and the coupling constants. 

If the seed and pump fields are shot-noise limited, $\tilde{V}_{A_{in}}^{1,2} = \tilde{V}_{B_{in}}^{1,2} = 1$, and the quadrature variances reduce to
\begin{eqnarray}
\tilde{V}_{A_{out}}^1(\Omega) &=& 
\sum^{4}_{j=1}\left(|\Theta^{(1j)}_{in}(\Omega)|^2 + |\Theta^{(1j)}_{out}(\Omega)|^2\right. \nonumber\\
&&+ \left.
|\Theta^{(1j)}_{sc}(\Omega)|^2 + 
|\Theta^{(1j)}_{abs}(\Omega)|^2\right),\\
\tilde{V}_{A_{out}}^2(\Omega) &=& 
\sum^{4}_{j=1}\left(|\Theta^{(2j)}_{in}(\Omega)|^2 + |\Theta^{(2j)}_{out}(\Omega)|^2\right. \nonumber\\
&&+ \left.
|\Theta^{(2j)}_{sc}(\Omega)|^2 + 
|\Theta^{(2j)}_{abs}(\Omega)|^2\right).
\end{eqnarray}

We now turn to the discussion of a limiting case to obtain simple analytic forms by making the following assumptions that are applicable to most practical cases of squeezing: (1) the pump noise is comparable to the seed noise, (2) the fundamental intra-cavity field is much weaker than the second-harmonic intra-cavity field ($|a|\ll|b|$), (3) there is no intra-cavity scattering, (4) there are no average cavity detunings at both the fundamental and second-harmonic frequencies, (5) the frequency of interest is within the linewidth of the OPA cavity, and (6) we are only interested in maximum phase squeezing ($\phi_b=0$) or maximum amplitude squeezing ($\phi_b=\pi$) so that any term proportional to $\bar{b}-\bar{b}^*$ is zero. Under these assumptions, we can make the following approximations: 
\widetext
For the amplitude quadrature variance,
\begin{eqnarray}
\label{eq:limiting_amplitude}
\Theta_{in}^{(11)} &\simeq& \frac{\sqrt{\gamma_a^{in}\gamma_a^{out}}(2\gamma_a^{tot}+\bar{\epsilon}\bar{b}^*+\bar{\epsilon}^*\bar{b})}{\gamma_a^{tot 2}-|\bar{\epsilon}|^2|\bar{b}|^2}, \hspace{3.6cm}
\Theta_{in}^{(12)} \simeq \frac{i\sqrt{\gamma_a^{in}\gamma_a^{out}}(\bar{\epsilon}^*\bar{b}-\bar{\epsilon}\bar{b}^*)}{\gamma_a^{tot 2}-|\bar{\epsilon}|^2|\bar{b}|^2},\nonumber\\
\Theta_{in}^{(13)} &\simeq& -\frac{1}{\Omega-i\Omega_T} \frac{2\sqrt{2}i\sqrt{\gamma_a^{out}\gamma_b^{abs}\gamma_b^{in}}\left[\bar{a}\bar{b}^*C_b(\gamma_a^{tot}+\bar{\epsilon}^*\bar{b})+\bar{a}^*\bar{b}C_b^*(\gamma_a^{tot}+\bar{\epsilon}\bar{b}^*)\right]}{\gamma_b^{tot}(\gamma_a^{tot 2}-|\bar{\epsilon}|^2|\bar{b}|^2)}, \hspace{2cm}
\Theta_{in}^{(14)} \simeq 0,\nonumber\\
\Theta_{out}^{(11)} &\simeq& \frac{-\gamma_a^{tot 2}+2\gamma_a^{out}\gamma_a^{tot}+\gamma_a^{out}(\bar{\epsilon}\bar{b}^*+\bar{\epsilon}^*\bar{b})+|\bar{\epsilon}|^2|\bar{b}|^2}{\gamma_a^{tot 2}-|\bar{\epsilon}|^2|\bar{b}|^2}, \hspace{1cm}
\Theta_{out}^{(12)} \simeq \frac{i\gamma_a^{out}(\bar{\epsilon}^*\bar{b}-\bar{\epsilon}\bar{b}^*)}{\gamma_a^{tot 2}-|\bar{\epsilon}|^2|\bar{b}|^2},\nonumber\\
\Theta_{out}^{(13)} &\simeq& -\frac{1}{\Omega-i\Omega_T} \frac{2\sqrt{2}i\sqrt{\gamma_a^{out}\gamma_b^{abs}\gamma_b^{out}}\left[\bar{a}\bar{b}^*C_b(\gamma_a^{tot}+\bar{\epsilon}^*\bar{b})+\bar{a}^*\bar{b}C_b^*(\gamma_a^{tot}+\bar{\epsilon}\bar{b}^*)\right]}{\gamma_b^{tot}(\gamma_a^{tot 2}-|\bar{\epsilon}|^2|\bar{b}|^2)}, \hspace{1.85cm}
\Theta_{out}^{(14)} \simeq 0,\nonumber\\
\Theta_{abs}^{(11)} &\simeq& \frac{\sqrt{\gamma_a^{abs}\gamma_a^{out}}(2\gamma_a^{tot}+\bar{\epsilon}\bar{b}^*+\bar{\epsilon}^*\bar{b})}{\gamma_a^{tot 2}-|\bar{\epsilon}|^2|\bar{b}|^2}, \hspace{3.5cm}
\Theta_{abs}^{(12)} \simeq \frac{i\sqrt{\gamma_a^{abs}\gamma_a^{out}}(\bar{\epsilon}^*\bar{b}-\bar{\epsilon}\bar{b}^*)}{\gamma_a^{tot 2}-|\bar{\epsilon}|^2|\bar{b}|^2},\nonumber\\
\Theta_{abs}^{(13)} &\simeq& -\frac{1}{\Omega-i\Omega_T} \frac{i\sqrt{2\gamma_a^{out}}(2\gamma_b^{abs}-\gamma_b^{tot})\left[\bar{a}\bar{b}^*C_b(\gamma_a^{tot}+\bar{\epsilon}^*\bar{b})+\bar{a}^*\bar{b}C_b^*(\gamma_a^{tot}+\bar{\epsilon}\bar{b}^*)\right]}{\gamma_b^{tot}(\gamma_a^{tot 2}-|\bar{\epsilon}|^2|\bar{b}|^2)}, \hspace{1.5cm}
\Theta_{abs}^{(14)} \simeq 0,\nonumber\\
\Theta_{sc}^{(11)} &\simeq& 0, \hspace{2cm}
\Theta_{sc}^{(12)} \simeq 0, \hspace{2cm}
\Theta_{sc}^{(13)} \simeq0, \hspace{2cm}
\Theta_{sc}^{(14)} \simeq 0, 
\end{eqnarray}
\endwidetext
\widetext
For the phase quadrature variance,
\begin{eqnarray}
\label{eq:limiting_phase}
\Theta_{in}^{(21)} &\simeq& \frac{i\sqrt{\gamma_a^{in}\gamma_a^{out}}(\bar{\epsilon}^*\bar{b}-\bar{\epsilon}\bar{b}^*)}{\gamma_a^{tot 2}-|\bar{\epsilon}|^2|\bar{b}|^2},\hspace{1.2cm}
\Theta_{in}^{(22)} \simeq \frac{\sqrt{\gamma_a^{in}\gamma_a^{out}}(2\gamma_a^{tot}-\bar{\epsilon}\bar{b}^*-\bar{\epsilon}^*\bar{b})}{\gamma_a^{tot 2}-|\bar{\epsilon}|^2|\bar{b}|^2}, \nonumber\\
\Theta_{in}^{(23)} &\simeq& -\frac{1}{\Omega-i\Omega_T} \frac{2\sqrt{2}\sqrt{\gamma_a^{out}\gamma_b^{abs}\gamma_b^{in}}\left[\bar{a}\bar{b}^*C_b(\gamma_a^{tot}-\bar{\epsilon}^*\bar{b})-\bar{a}^*\bar{b}C_b^*(\gamma_a^{tot}-\bar{\epsilon}\bar{b}^*)\right]}{\gamma_b^{tot}(\gamma_a^{tot 2}-|\bar{\epsilon}|^2|\bar{b}|^2)}, \hspace{1.86cm}
\Theta_{in}^{(24)} \simeq 0,\nonumber\\
\Theta_{out}^{(21)} &\simeq& \frac{i\gamma_a^{out}(\bar{\epsilon}^*\bar{b}-\bar{\epsilon}\bar{b}^*)}{\gamma_a^{tot 2}-|\bar{\epsilon}|^2|\bar{b}|^2}, \hspace{2cm}
\Theta_{out}^{(22)} \simeq \frac{-\gamma_a^{tot 2}+2\gamma_a^{out}\gamma_a^{tot}-\gamma_a^{out}(\bar{\epsilon}\bar{b}^*+\bar{\epsilon}^*\bar{b})+|\bar{\epsilon}|^2|\bar{b}|^2}{\gamma_a^{tot 2}-|\bar{\epsilon}|^2|\bar{b}|^2},\nonumber\\
\Theta_{out}^{(23)} &\simeq& -\frac{1}{\Omega-i\Omega_T} \frac{2\sqrt{2}\sqrt{\gamma_a^{out}\gamma_b^{abs}\gamma_b^{out}}\left[\bar{a}\bar{b}^*C_b(\gamma_a^{tot}-\bar{\epsilon}^*\bar{b})-\bar{a}^*\bar{b}C_b^*(\gamma_a^{tot}-\bar{\epsilon}\bar{b}^*)\right]}{\gamma_b^{tot}(\gamma_a^{tot 2}-|\bar{\epsilon}|^2|\bar{b}|^2)}, \hspace{1.75cm}
\Theta_{out}^{(24)} \simeq 0,\nonumber\\
\Theta_{abs}^{(21)} &\simeq& \frac{i\sqrt{\gamma_a^{abs}\gamma_a^{out}}(\bar{\epsilon}^*\bar{b}-\bar{\epsilon}\bar{b}^*)}{\gamma_a^{tot 2}-|\bar{\epsilon}|^2|\bar{b}|^2}, \hspace{1.04cm}
\Theta_{abs}^{(22)} \simeq \frac{\sqrt{\gamma_a^{abs}\gamma_a^{out}}(2\gamma_a^{tot}-\bar{\epsilon}\bar{b}^*-\bar{\epsilon}^*\bar{b})}{\gamma_a^{tot 2}-|\bar{\epsilon}|^2|\bar{b}|^2}, \nonumber\\
\Theta_{abs}^{(23)} &\simeq& -\frac{1}{\Omega-i\Omega_T} \frac{\sqrt{2\gamma_a^{out}}(2\gamma_b^{abs}-\gamma_b^{tot})\left[\bar{a}\bar{b}^*C_b(\gamma_a^{tot}-\bar{\epsilon}^*\bar{b})-\bar{a}^*\bar{b}C_b^*(\gamma_a^{tot}-\bar{\epsilon}\bar{b}^*)\right]}{\gamma_b^{tot}(\gamma_a^{tot 2}-|\bar{\epsilon}|^2|\bar{b}|^2)}, \hspace{1.4cm}
\Theta_{abs}^{(24)} \simeq 0,\nonumber\\
\Theta_{sc}^{(21)} &\simeq& 0, \hspace{2cm}
\Theta_{sc}^{(22)} \simeq 0, \hspace{2cm}
\Theta_{sc}^{(23)} \simeq0, \hspace{2cm}
\Theta_{sc}^{(24)} \simeq 0.
\end{eqnarray}
\endwidetext
Note that in this limiting case, the noise coupling constants, $\Theta_{in}^{13}, \Theta_{out}^{13}, \Theta_{abs}^{13}, \Theta_{in}^{23}, \Theta_{out}^{23}$, and $\Theta_{abs}^{23}$, have the frequency dependence of $1/(\Omega-i\Omega_T)$, and therefore they increase as the frequency decreases for $\Omega \gg \Omega_T$, degrading the squeezing level at low frequencies. In Section~\ref{sect:results}, we will discuss which couping constants are dominant at high frequencies, low frequencies, and intermediate frequencies. 

\section{Results}
\label{sect:results}
In this section, we discuss various cases of the influence of the photothermal noise on squeezed quadrature variances in both amplitude and phase quadratures. The most significant photothermal effects are seen in the phase quadrature, and therefore we mainly discuss quadrature variances in the phase quadrature. Section~\ref{sect:noise_variances} presents such results. Section~\ref{sect:GW} discusses the effect of squeezing with the photothermal noise on a conventional gravitational wave interferometer when a photothermal-noise-limited squeezed field is injected into it. The following plots are obtained from the exact normalized quadrature variances in Eqs.~\eqref{eq:V1_Aout} and~\eqref{eq:V2_Aout} with realistic values for OPA parameters, which are listed in Table~\ref{fig:Table I}. The effect of green-induced infrared absorption (GRIIRA)~\cite{furukawa} is not considered in this paper.

\subsection{Normalized Quadrature Variances with the Photothermal Effect}
\label{sect:noise_variances}
\begin{figure}[t]
\includegraphics[width=0.45\textwidth]{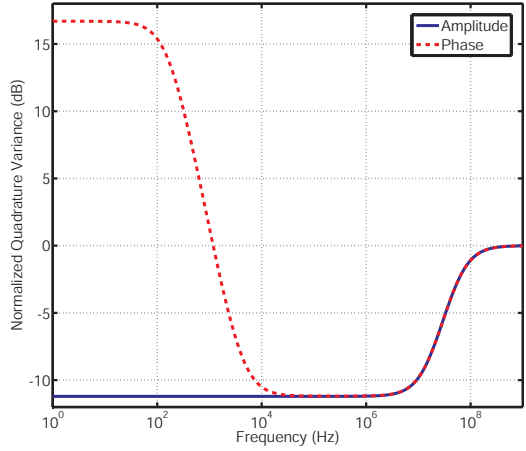}
\caption{\label{fig:fig2} The comparison of normalized amplitude and phase quadrature variances relative to the shot noise vs frequency with the photothermal effect. The amplitude quadrature is relatively immune to the photothermal noise in most practical cases. The seed power is 1 mW. The pump power is 0.5P$_{th}$. The input fields are shot noise limited.}
\end{figure}

The amplitude quadrature is relatively immune to the photothermal noise for the following reasons. The intra-cavity fundamental field is deamplified in the degenerate parametric oscillation, and thus the noise coupling is smaller than in the phase squeezing case. Moreover, in the ideal case, the system is held on resonance and operated at the phase-matched temperature. The detuning fluctuations do not couple into the amplitude quadrature for a cavity on resonance since the frequency derivative of the amplitude response of the cavity is zero. Figure~\ref{fig:fig2} compares the effect of the photothermal noise between the amplitude and phase squeezing cases. (Note that they are not obtained simultaneously from the OPA; different pump phases are required.) The photothermal noise is therefore significant in the phase quadrature and relatively unimportant in the amplitude quadrature, in most practical cases. In the absence of the photothermal noise, the normalized quadrature variance would be flat within the OPA linewidth.

\begin{figure}[t]
\includegraphics[width=0.45\textwidth]{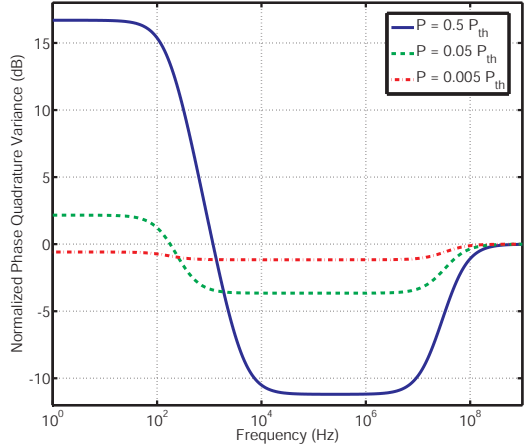}
\caption{\label{fig:fig3} The normalized phase quadrature variance relative to the shot noise vs frequency with the photothermal effect for different pump powers. The photothermal cutoff frequency is higher for a higher level of squeezing whereas it is lower for a lower level of squeezing. The seed power is 1 mW. The input fields are shot noise limited.}
\end{figure}

\begin{figure}[t]
\includegraphics[width=0.45\textwidth]{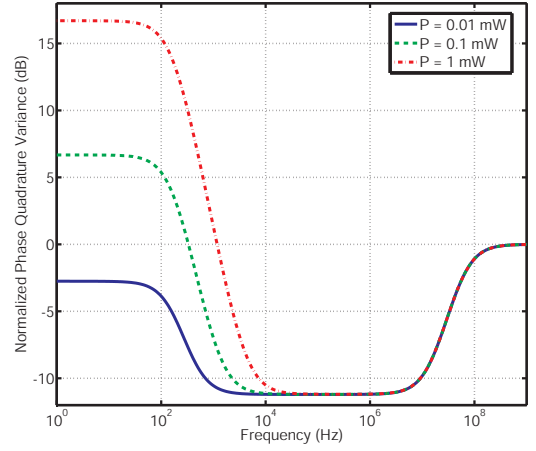}
\caption{\label{fig:fig4} The normalized phase quadrature variance relative to the shot noise vs frequency with the photothermal effect for different seed powers. The photothermal cutoff frequency is higher for a higher seed power. The pump power is 0.5P$_{th}$. The input fields are shot noise limited.}
\end{figure}

\begin{figure}[t]
\includegraphics[width=0.45\textwidth]{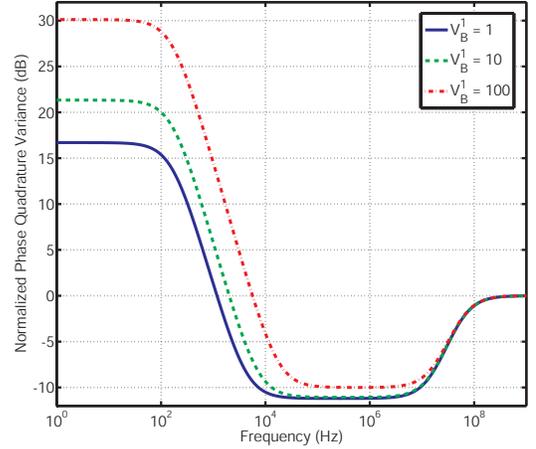}
\caption{\label{fig:fig5} The normalized phase quadrature variance relative to the shot noise vs frequency with the photothermal effect for different pump noise levels. The photothermal cutoff frequency is higher for a higher pump noise level. The seed power is 20 mW. The pump power is 0.5P$_{th}$.}
\end{figure}

As can be seen in Figures~\ref{fig:fig2},~\ref{fig:fig3},~\ref{fig:fig4}, and~\ref{fig:fig5} in which the normalized quadrature variances versus frequency are plotted, the squeezing level is cut off at frequencies below 10 kHz (depending on the OPA cavity parameters) due to the photothermal noise which has $1/(\Omega^2+\Omega_T^2)$ roll-off in variance. Parameter values used for the figures are summarized in Table~\ref{fig:Table I}. The high frequency cutoff is due to the linewidth of the OPA cavity below which the seed field is squeezed. At frequencies above the high cutoff frequency, $\Theta^{(11)}_{out}$ and $\Theta^{(21)}_{out}$ dominate in the amplitude and phase quadratures respectively. At frequencies between the two cutoff frequencies, $\Theta^{(11)}_{in}, \Theta^{(11)}_{out}, \Theta^{(11)}_{abs}, \Theta^{(22)}_{in}, \Theta^{(22)}_{out}$, and $\Theta^{(22)}_{abs}$ dominate (depending on the OPA cavity parameters) in the amplitude and phase quadratures respectively. The photothermal cutoff frequency is greater than the adiabatic limit $\Omega_T$ in most practical cases, and therefore, at low frequencies above $\Omega_T$, $\Theta^{(13)}_{in}, \Theta^{(13)}_{abs}, \Theta^{(23)}_{in}$, and $\Theta^{(23)}_{abs}$ that have the frequency dependence of $1/\Omega$ dominate (depending on the OPA cavity parameters) in the amplitude and phase quadratures respectively, as can easily be seen in Eqs.~\eqref{eq:limiting_amplitude} and~\eqref{eq:limiting_phase}. At frequencies below the adiabatic limit $\Omega_T$, the quadrature variances become flat. The domination of these $\Theta$'s in each frequency band is valid regardless of which quadrature variance is squeezed/anti-squeezed. 

\begin{figure*}[bth]
\begin{center}
\begin{tabular}{cc}
\includegraphics[width=0.45\textwidth]{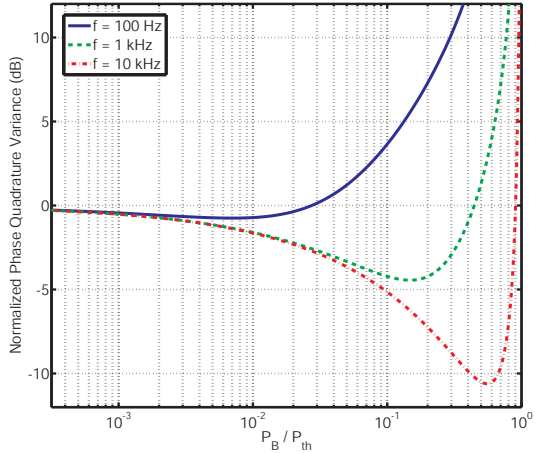}
\hspace{0.025\textwidth} & \hspace{0.025\textwidth}
\includegraphics[width=0.45\textwidth]{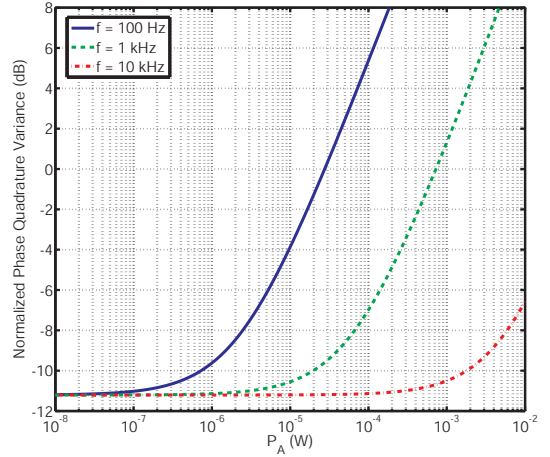}
\end{tabular}
\end{center}
\caption{\textit{Left}: The normalized phase quadrature variance relative to the shot noise level vs pump power with the photothermal effect for different frequencies. The seed power is 1 mW. \textit{Right}: The normalized phase quadrature variance relative to the shot noise level vs seed power with the photothermal effect for different frequencies. The pump power is 0.5P$_{th}$. The input fields are shot noise limited in both graphs.}
\label{fig:fig6}
\end{figure*}

Figure~\ref{fig:fig3} shows the normalized quadrature variance of a phase squeezed state as a function of frequency for various pump powers. As the pump power approaches the OPA threshold, the squeezing level at high frequencies increases, but higher pump power also increases the photothermal noise contribution. The photothermal noise is largest at low frequencies and limits squeezing to occur only at higher frequencies where the photothermal noise is small. The increase in the photothermal noise as the pump power approaches the OPA threshold is also attributable to the increase in the fundamental field amplitude via high parametric gain which increases the photothermal noise coupling. Squeezing at lower frequencies can be acquired at the expense of the broadband level of squeezing. 

Figure~\ref{fig:fig4} shows the normalized quadrature variance of a phase squeezed state as a function of frequency for various seed powers. Since the coupling of the photothermal effect to the quadrature variances is proportional to the seed power, the photothermal noise contribution limits squeezing to higher frequencies as the seed power increases. Hence, the seed power should be set as small as possible to avoid the photothermal noise. However, the reduction of the seed power leads to difficulties in obtaining an optical signal for controlling the phase of the seed. Therefore, in practice, the seed power should be properly chosen such that it optimizes control stability and the frequency of interest is above the photothermal cutoff frerequency. A crystal with a smaller absorption rate can also reduce the pump noise coupling and therefore the photothermal noise. Its effect appears similar to the effect of a lower seed power as shown in Figure~\ref{fig:fig4}. 

Figure~\ref{fig:fig5} shows the normalized quadrature variance of a phase squeezed state as a function of frequency for various pump amplitude noise levels. As the pump noise increases, the overall squeezing level also decreases due to the direct coupling of the pump noise to the quadrature variances~\cite{wodkiewicz,crouch,zubairy}. At the same time, the photothermal noise induced by the pump noise also increases, driving up the photothermal noise limited frequency. 
\begin{figure}[t]
\includegraphics[width=0.45\textwidth]{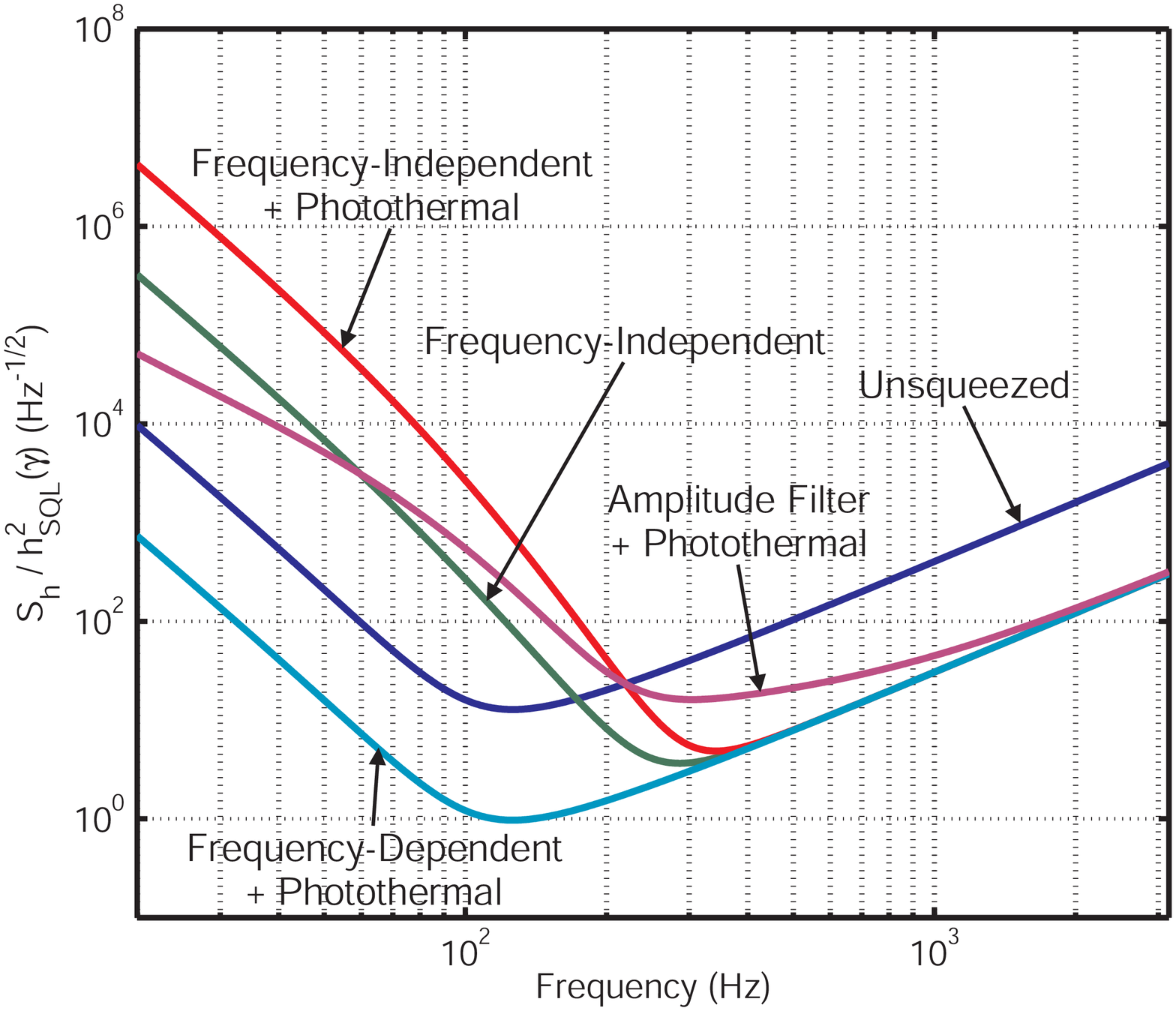}
\caption{\label{fig:fig7} The spectral noise density, normalized by the standard quantum limit (SQL), for a conventional GW interferometer with (i) no squeezed input (Unsqueezed), (ii) squeezed light injected without the photothermal noise (Frequency-Independent), (iii) squeezed light injected with a frequency-dependent squeeze angle and the photothermal noise (Frequency-Dependent + Photothermal), (iv) squeezed light injected with the photothermal noise (Frequency-Independent + Photothermal), and (v) amplitude-filtered squeezed light injected with the photothermal noise (Amplitude Filter + Photothermal). The filter linewidth is $2\pi\times 400$ Hz. The input squeezed source is chosen to be photothermal-noise-limited amplitude squeezed light since the amplitude quadrature is relatively immune to the photothermal noise. The seed and pump powers are 10 mW and 0.5P$_{th}$ respectively. The seed and pump fields are shot noise limited. The anti-squeezed phase quadrature variance has the photothermal cutoff frequency at about 1 kHz and the adiabatic limit at about 100 Hz.}
\end{figure}
\begin{center}
\begin{table*}[t]
\caption{\label{fig:Table I} OPA Cavity Parameters}
\begin{center}
\begin{tabular}{|c|c|c|c|}
\hline
Parameter & Symbol & Value & Units\\
\hline
Fundamental Wavelength & $\lambda_a$ & 1064 & nm \\
Second-Harmonic Wavelength & $\lambda_b$ & 532 & nm \\
Reflectivity of Input Coupler at Fundamental Frequency & $R_a^{in}$ & 99.96 & $\%$ \\
Reflectivity of Output Coupler at Fundamental Frequency & $R_a^{out}$ & 95.6 & $\%$ \\
Reflectivity of Input Coupler at Second-Harmonic Frequency & $R_b^{in}$ & 4.0 & $\%$ \\
Reflectivity of Output Coupler at Second-Harmonic Frequency & $R_b^{out}$ & 99.96 & $\%$ \\
Absorption Rate at Fundamental Frequency & $\sigma_a^{abs}$ & 0.1 & $\%/\mbox{cm}$ \\
Scattering Rate at Fundamental Frequency & $\sigma_a^{sc}$ & 0.02 & $\%/\mbox{cm}$ \\
Absorption Rate at Second-Harmonic Frequency & $\sigma_b^{abs}$ & 4.0 & $\%/\mbox{cm}$ \\
Scattering Rate at Second-Harmonic Frequency & $\sigma_b^{sc}$ & 0.5 & $\%/\mbox{cm}$ \\
Crystal Length & $z$ & 7.5 & mm \\ 
Nonlinear Coupling Strength & $\kappa_0$ & 800,000 & 1/m/s \\
Phase-Matched Refractive Index & n & 2.233 & - \\
Specific Heat of Crystal & C & 633 & J/kg/K \\
Density of Crystal & $\rho$ & 4.648 & g/cm$^{3}$ \\
Thermal Conductivity of Crystal & $\kappa$ & 4 & W/K/m \\
Radius of Nonlinear Interaction & $r_0$ & 36 & $\mu$m \\
Phase Mismatch Constant & $\xi$ & 749 & 1/m/K \\
Thermal Expansion Constant in Ordinary Axis & $\alpha_a$ & $5\times 10^{-6}$ & 1/K \\
Thermal Expansion Constant in Extraordinary Axis & $\alpha_b$ & $5\times 10^{-6}$ & 1/K \\
Photo-refractive Constant in Ordinary Axis & $dn_a/dT$ & $3.3\times 10^{-6}$ & 1/K\\
Photo-refractive Constant in Extraordinary Axis & $dn_b/dT$ & $37.0\times 10^{-6}$ & 1/K\\
Temperature Offset & $\Delta T$ & 0.001 & K \\
Cavity Detuning at Fundamental Frequency & $\omega_a^{det}$ & 0 & Hz \\
Cavity Detuning at Second-Harmonic Frequency & $\omega_b^{det}$ & 0 & Hz \\
\hline
\end{tabular}
\end{center}
\label{default}
\end{table*}
\end{center}

Figure~\ref{fig:fig6} shows the quadrature variance of a phase squeezed state as a function of pump/seed power for various frequencies. Without the photothermal effect, the maximum squeezing would be achieved at the OPA threshold. In the presence of the photothermal noise, the squeezing level starts to degrade as the pump power approaches the OPA threshold. The photothermal noise can also be minimized by reducing the seed power. 

In summary, the undesirable consequences of the photothermal noise can be minimized by satisfiying the following conditions: (1) a low seed power, (2) a quiet pump field, (3) a crystal with a low absorption rate, and (4) squeeze the amplitude quadrature variance rather than the phase quadrature variance. 

\subsection{The Effect of the Photothermal Noise on Gravitational Wave Interferometers}
\label{sect:GW}
One primary purpose of low frequency squeezing is to improve the sensitivity of GW interferometers in the GW band which is typically 10 - 10,000 Hz~\cite{caves}. To implement it, a low-frequency squeezed field needs to be prepared for injection to the dark port of the GW interferometers. However, as discussed in Section~\ref{sect:noise_variances}, the squeezing level of phase-squeezed light is limited by the photothermal noise at low frequencies, and hence it places an important limit on the use of squeezed light in the GW interferometers. 

For a conventional GW interferometer with arm lengths $L$ and mirror masses $m$, the spectral density of the GW noise when a realistic squeezed field is injected to the dark port is given by~\cite{kimble}
\begin{eqnarray}
\label{eq:GW_IFO_sensitivity}
\hspace{-0.5cm}
& &\hspace{-1cm} \tilde{S}(\Omega) = \nonumber\\
& &\hspace{-1cm} \frac{h_{SQL}^2}{2}\left(\frac{1}{\eta}+\eta\right) \left(\tilde{V}_A^1\cos^2(\theta+\Phi)+\tilde{V}_A^2\sin^2(\theta+\Phi)\right),
\end{eqnarray}
where
\begin{eqnarray}
h_{SQL}(\Omega) \equiv \sqrt{\frac{8\hbar}{m\Omega^2 L^2}}
\end{eqnarray}
is the noise spectral density of the dimensionless GW strain at the standard quantum limit (SQL) for a GW interferometer with uncorrelated radiation pressure noise and shot noise, 
\begin{eqnarray}
\eta(\Omega) = \frac{2(I_0/I_{SQL})\gamma^4}{\Omega^2(\gamma^2+\Omega^2)}
\end{eqnarray}
is the effective coupling constant that relates the output signal to the motion of the GW interferometer mirrors, $\tilde{V}_A^1(\Omega), \tilde{V}_A^2(\Omega)$ are the amplitude and phase quadrature variances of the input squeezed field with a squeeze angle $\theta(\Omega)$ respectively, and
\begin{eqnarray}
\Phi(\Omega) &\equiv& \cot^{-1}\eta(\Omega) \nonumber\\
&=& \cot^{-1}\left[\frac{2(I_0/I_{SQL})\gamma^4}{\Omega^2(\gamma^2+\Omega^2)}\right]. 
\end{eqnarray}
Here $\gamma$ is the linewidth of the arm cavities, $I_0$ is the optical power to the beam-splitter of the GW interferometer, and $I_{SQL}$ is the optical power to reach the SQL. 

Figure~\ref{fig:fig7} shows the noise spectral density for the conventional GW interferometer in various cases when a photothermal-noise-limited amplitude-squeezed field is injected into the interferometer. The effect of squeezing with the photothermal noise on the sensitivity of the GW interferometer is plotted in the figure. Since the amplitude quadrature is relatively insensitive to the photothermal noise, we choose the squeeze angle $\theta$ such that the amplitude quadrature variance is squeezed and the phase quadrature variance is anti-squeezed. The seed and pump powers to the OPA cavity are 10 mW and 0.5P$_{th}$. If such a squeezed light field is used for injection without a rotation of the squeeze angle, it degrades the spectral noise density at about 200 Hz compared with the unsqueezed case although it reduces shot noise at high frequencies. If a set of two filter cavities is used to give the frequency dependent squeeze angle $\theta(\Omega)$ such that $\theta(\Omega) = -\Phi(\Omega)$~\cite{kimble}, the sensitivity is improved at all frequencies. This is because the second term in Eq.~\eqref{eq:GW_IFO_sensitivity} becomes zero and the phase quadrature variance does not couple into the sensitivity curve. However, it is likely to be difficult and costly to implement such a squeeze angle rotation since the length of the filter cavities is required to be long (on the order of kilometers) in order to minimize losses that destroy squeezing in the process~\cite{kimble,harms}. If a squeeze amplitude filter is instead used before injecting the squeezed light into the GW interferometer~\cite{corbitt} such that for a filter linewidth $\gamma_f$,
\begin{eqnarray}
\tilde{S}(\Omega) = \frac{h_{SQL}^2}{2\eta}\left[\zeta_1(\tilde{V}_1+\eta^2 \tilde{V}_2)+\zeta_2(1+\eta^2)\right],
\end{eqnarray}
where 
\begin{eqnarray}
\zeta_1(\Omega) = \frac{\Omega^2}{\gamma_f^2+\Omega^2}, 
\hspace{0.5cm}
\zeta_2(\Omega) = \frac{\gamma_f^2}{\gamma_f^2+\Omega^2},
\end{eqnarray}
the phase quadrature variance containing the photothermal noise can be reduced at frequencies below 200 Hz although the level of squeezing at frequencies above 200 Hz is slightly decreased. Here $\gamma_f = 2\pi\times 400$ Hz is used. Note that a photothermal-noise-limited phase-squeezed input field with similar experimental parameters (the seed power = 1 mW, the pump power = 0.5P$_{th}$) does not enable quantum noise reduction below 400 Hz, even if the optimal frequency dependent squeeze angle rotation is applied. 

\section{Conclusions}
\label{sect:conclusions}
We have derived and solved the field evolution equations in the degenerate optical parametric amplifier (OPA) with the photothermal noise through the photo-refractive effect and thermal expansion of nonlinear crystals. We also have discussed various cases about the effect of the photothermal noise on amplitude and phase quadrature variances. We have found that the photothermal noise in the OPA introduces a significant amount of noise on phase squeezed beams, making them less than ideal for low-frequency applications such as GW interferometers, whereas amplitude squeezed beams are less sensitive to the photothermal noise and may provide a better choice for low frequency applications. This problem can be solved by reducing the seed power and pump noise and using a nonlinear crystal with a low absorption rate in order to decrease the photothermal noise.   

\section{acknowledgments}
\label{sect:acknowledgments}
We would like to thank our colleagues and collaborators at the LIGO Laboratory and Center for Gravitational Physics at the Australian National University, especially Thomas Corbitt, David Ottaway, Stanley Whitcomb, and Kentaro Somiya for valuable discussions. We also thank our colleague, Sergey P. Vyatchanin at Moscow State University for correcting an error. We gratefully acknowledge support from National Science Foundation grants PHY-0107417 and PHY-0300345 and the Australian Research Council.



\end{document}